\documentclass[11pt]{article}
\usepackage{subfigure}
\usepackage{graphicx}
\usepackage{hyperref}
\usepackage{epsfig,subfigure}
\usepackage{color}
\usepackage{enumerate}

\usepackage{float}

\newcommand{\define}{\stackrel{\triangle}{=}}
\def\QED{\mbox{\rule[0pt]{1.5ex}{1.5ex}}}
\def\proof{\noindent\hspace{2em}{\it Proof: }}

\makeatletter
\def\bbordermatrix#1{\begingroup \m@th
  \@tempdima 4.75\p@
  \setbox\z@\vbox{%
    \def\cr{\crcr\noalign{\kern2\p@\global\let\cr\endline}}%
    \ialign{$##$\hfil\kern2\p@\kern\@tempdima&\thinspace\hfil$##$\hfil
      &&\quad\hfil$##$\hfil\crcr
      \omit\strut\hfil\crcr\noalign{\kern-\baselineskip}%
      #1\crcr\omit\strut\cr}}%
  \setbox\tw@\vbox{\unvcopy\z@\global\setbox\@ne\lastbox}%
  \setbox\tw@\hbox{\unhbox\@ne\unskip\global\setbox\@ne\lastbox}%
  \setbox\tw@\hbox{$\kern\wd\@ne\kern-\@tempdima\left[\kern-\wd\@ne
    \global\setbox\@ne\vbox{\box\@ne\kern2\p@}%
    \vcenter{\kern-\ht\@ne\unvbox\z@\kern-\baselineskip}\,\right]$}%
  \null\;\vbox{\kern\ht\@ne\box\tw@}\endgroup}
\makeatother

\usepackage{arydshln}
\usepackage{amsmath,amsfonts,amssymb}
\usepackage{verbatim}
\usepackage[bookmarks=false]{}
\usepackage[font={small}]{caption}

\newtheorem{theorem}{\bf Theorem}

\newtheorem{lemma}{Lemma}

\newcommand\blfootnote[1]{%
  \begingroup
  \renewcommand\thefootnote{}\footnote{#1}%
  \addtocounter{footnote}{-1}%
  \endgroup
}

\begin{document}
\date{}
%

\title{Elevated Multiplexing and Signal Space Partitioning in the 2 User MIMO IC with Partial CSIT}
\author{Bofeng Yuan and Syed A. Jafar\\
Center for Pervasive Communications and Computing (CPCC)\\
University of California, Irvine\\
Email: $\left\{\right.$bofengy, syed$\left.\right\}$@uci.edu}




\maketitle
\blfootnote{Submitted on March 18, 2016 as an invited paper to IEEE SPAWC. Bofeng Yuan (email: bofengy@uci.edu) and Syed A. Jafar (email: syed@uci.edu) are with the Center of Pervasive Communications and Computing (CPCC) in the Department of Electrical Engineering and Computer Science (EECS) at the University of California Irvine.  }

\begin{abstract}
The $2$ user MIMO interference channel with arbitrary antenna configurations is studied under arbitrary levels of partial CSIT for each of the channels, to find the degrees of freedom (DoF) achievable by either user while the other user achieves his full interference-free DoF. The goal is to gain new insights due to the inclusion of  MIMO (multiple antennas at both transmitters and receivers) into the signal space partitioning schemes  associated with partial CSIT.  An interesting idea that emerges from this study is ``elevated multiplexing" where the signals are split into  streams and transmitted from separate antennas at elevated power levels, which allows these signals to be jointly decoded at one receiver which has fewer spatial dimensions  with lower interference floors, while another receiver is simultaneously able to separately decode these signals with a higher interference floor but across a greater number of spatial dimensions. Remarkably, we find that there is a DoF benefit from increasing the number of  antennas at a transmitter even if that transmitter already has more antennas than its desired receiver and has no CSIT. 
\end{abstract}

\newpage
\section{Introduction}
Degrees of freedom studies of wireless networks have contributed many fundamental insights into their capacity limits \cite{Jafar_FnT}. One of the most critical determinants of these capacity limits is the amount of channel state information at the transmitters (CSIT).
\subsection*{Sharp Contrast between Perfect and Finite Precision CSIT}
 At one extreme, if the CSIT is perfect, i.e., available with infinite precision, then tremendous DoF gains are possible, mainly through zero-forcing and interference alignment \cite{Jafar_FnT}. At the other extreme, if the CSIT is absent then the DoF collapse \cite{Huang_Jafar_Shamai_Vishwanath, Vaze_Varanasi_MIMOBC_Delayed, Zhu_Guo_MIMOIC,Lozano_Heath_Andrews}. In fact, even if partial CSIT is present, as long as it is limited to finite precision, then the DoF still collapse \cite{Arash_Jafar_GC14}. For example, consider an arbitrary channel coefficient $H_{ij}$, which is modeled under partial CSIT as 
 \begin{eqnarray*}
 H_{ij}&=&\hat{H}_{ij}+\sqrt{\epsilon}\tilde{H}_{ij}
 \end{eqnarray*}
 so that $\hat{H}_{ij}$ is the channel estimate known to the transmitter, while $\tilde{H}_{ij}$ is the normalized estimation error, with mean squared error  $\epsilon>0$. Even if $\epsilon$ is very small, as long as $\epsilon$ does not diminish with SNR $(P)$,\footnote{Following convention, we use $P$ to represent the nominal SNR variable.} the DoF collapse. Since in practice, CSIT can only be obtained to finite precision, at first sight the collapse of DoF under finite precision CSIT seems to suggest that there is no benefit of zero-forcing or interference alignment techniques in practice. 
\subsection*{Expanding the DoF Formulation to Capture Partial CSIT} 
Upon careful assessment it becomes evident that the collapse of DoF under finite precision CSIT is primarily due to the limitation of the traditional DoF formulation which cannot distinguish between the relative strengths of constants (e.g., any non-zero channel, regardless of its strength, carries 1 DoF). Intuitively, from a DoF perspective a small estimation error $\epsilon$ is no different than a large estimation error. Since the quality of channel estimates is a such a crucial factor, it is important to expand the DoF formulation to be non-trivially responsive to this parameter. Motivated by the generalized degrees of freedom (GDoF) framework,  the channel estimation error strength is captured in the parameter $\beta$, so that $\epsilon=P^{-\beta}$ \cite{Sheng_Kobayashi_Gesbert_Yi, Gou_Jafar,Arash_Jafar_GC14, Joudeh_Clerckx, Arash_Jafar_ISIT16, Chenxi_Bruno}. With this formulation, it turns out that in the DoF sense, $\beta=1$ corresponds to perfect CSIT while $\beta=0$ corresponds to no CSIT (also finite precision CSIT). As  $\beta$ spans the range of  values between $0$ and $1$ it  captures all intermediate levels of partial CSIT. \emph{While the scaling of estimation error with SNR may seem unnatural for a given channel, the  interpretation consistent with the GDoF framework, is not that the SNR is increasing for a given channel, but rather that a given channel is only associated with a given SNR.} As SNR value is allowed to increase, each new value of SNR defines a new channel. The reason this class of channels is studied together is because, normalized by $\log(\mbox{SNR})$, they have the same approximate capacity. Indeed this is precisely how the GDoF metric has been used to find the approximate capacity of several wireless networks of interest including, most prominently, the capacity characterization of the 2 user interference channels to an accuracy of within 1 bit for all choices of channel parameters \cite{Etkin_Tse_Wang}.

\subsection*{DoF under Partial CSIT: Signal Space Partitioning}
DoF under partial CSIT have been studied under a variety of settings \cite{Sheng_Kobayashi_Gesbert_Yi, Gou_Jafar,Arash_Jafar_GC14}. A common observation repeatedly encountered in these studies is the idea of signal space partitioning in accordance with partial CSIT. Starting from the earliest instances in \cite{Gou_Jafar, Sheng_Kobayashi_Gesbert_Yi}, essentially the same phenomenon has been recognized independently as interference enhancement\footnote{As explicitly shown in \cite{Gou_Jafar} and also observed recently in \cite{ Arash_Jafar_ISIT16} for the vector broadcast setting, the partial CSIT setting framework where estimation error decays with a constant negative exponent of SNR, translates into the GDoF framework where  channels have  strengths that scale with different SNR exponents and only finite precision CSIT is available. This is because without loss of generality, the transmitter can rotate its signal space to map estimated zero-forcing directions directly to specific transmit antennas. As such GDoF studies under finite precision CSIT translate into DoF studies under partial CSIT.} \cite{Arash_Jafar_GC15} and topological rate-splitting \cite{Joudeh_Clerckx}. The broad implications of signal space partitioning are most recently highlighted in  \cite{Arash_Jafar_ISIT16} as follows. Essentially,  the signal space is partitioned according to the partial CSIT level $\beta$, so that  the bottom $\beta$ power levels correspond to perfect CSIT, while the remaining top $1-\beta$ power levels correspond to no CSIT. To understand the idea of signal space partitioning intuitively, consider a wireless network where all channels are subject to CSIT level $\beta$. For each transmitter in this network, the  transmit signal $X$ (subject to transmit power $P$) is decomposed into two parts $\hat{X}$ and $\tilde{X}$ corresponding to perfect and no CSIT respectively, each normalized to unit power, and each encoded  independently from  Gaussian codebooks, so that
\begin{eqnarray*}
X&=&\sqrt{P^\beta}\hat{X}+\sqrt{P-P^\beta}\tilde{X}\\
&=&\sqrt{P^\beta}\hat{X}+\sqrt{P}\tilde{X}+O(1)
\end{eqnarray*}
where $O(1)$ is a  negligible term for DoF purposes whose power is bounded by a constant.
Note that as this transmitted signal goes through a channel $H$, its contribution to the received signal is
\begin{eqnarray*}
HX&=&(\hat{H}+\sqrt{P^{-\beta}}\tilde{H})(\sqrt{P^\beta}\hat{X}+\sqrt{P}\tilde{X}+O(1))\\
&=&\sqrt{P^\beta}\hat{H}\hat{X}+\sqrt{P}{H}\tilde{X}+O(1)
\end{eqnarray*}
Thus, at each receiver, all the different $\tilde{X}$ signals from every transmitter are received at power $\sim P$, while the $\hat{X}$ signals from every transmitter are received at power $\sim P^{\beta}$. The $\tilde{X}$ signals go through the partially known channel $H$, and carry only common message(s) which are decoded by every receiver (e.g., as a multiple access channel) while treating the interference from the $\hat{X}$ parts as noise. Since this decoding has an SINR value $P/P^{\beta}=P^{1-\beta}$, the common messages achieve a total of $1-\beta$ DoF. Once the $\tilde{X}$ terms are decoded and subtracted out, only the $\hat{X}$ terms are left. For these terms note that the SNR is $P^\beta$ and very importantly, these terms only go through the channel estimate $\hat{H}$ which is perfectly known to the transmitter. Therefore, the $\hat{X}$ signals are able to achieve $\beta$ times the DoF value under perfect CSIT.

This achievability argument based on signal space partitioning is broadly applicable. For example, consider the $K$ user interference channel, which has $\hat{D}=K/2$ DoF under perfect CSIT \cite{Cadambe_Jafar_int} and only $\tilde{D}=1$ DoF under finite-precision CSIT \cite{Arash_Jafar_GC14}. If all channels have channel uncertainty level $\beta$, then the $K$ user interference channel achieves $\beta\hat{D}$ DoF from the $\hat{X}$ codewords and $(1-\beta)\tilde{D}$ DoF from the $\tilde{X}$ codewords, for a total of $\frac{K}{2}\beta+1-\beta$ DoF. Similarly, the $X$ channel which has $K_1$ transmitters and $K_2$ receivers, and achieves $\hat{D}=K_1K_2/(K_1+K_2-1)$ DoF under perfect CSIT, and only $\tilde{D}=1$ DoF under finite precision CSIT, achieves $\beta\hat{D}+(1-\beta)\tilde{D}$ with partial CSIT level $\beta$. As the final example, consider the MISO BC with $K$ transmit antennas and $K$ single antenna users which has $\hat{D}=K$ DoF with perfect CSIT and only $\tilde{D}=1$ DoF under finite precision CSIT. With partial CSIT level $\beta$, this channel has exactly $\beta\hat{D}+(1-\beta)\tilde{D}$ DoF as shown in  \cite{Arash_Jafar_ISIT16} where both achievability and outer bound are shown to prove the optimality of this DoF value, and therefore also the optimality of signal space partitioning under partial CSIT. Thus, note that as the partial CSIT level $\beta$ spans the range between $0$ and $1$, it bridges the contrasting extremes of DoF under perfect CSIT and finite precision CSIT. 

The idea of signal space partitioning for partial CSIT allows generalizations to  settings with asymmetric $\beta$ parameters through multilevel hierarchical partitions, with each power level (measured in terms of the exponent of $P$) allowing perfect CSIT for those links whose CSIT parameters $\beta$ are at that level or higher. Progress along these lines is reported in \cite{Chenxi_Bruno}. Another generalization, reported in \cite{Arash_Jafar_ISIT16}, explores the role of partial CSIT is conjunction with  the diversity of channel strengths as measured through power exponents in the GDoF framework. However, a most interesting direction that remains unexplored is the role of signal space partitioning in MIMO interference channels, especially with arbitrary antenna configurations and arbitrary partial CSIT levels. This is the direction that we wish to explore in this work. 

We explore the DoF of a 2 user MIMO interference channel with arbitrary antenna configurations ($M_1, M_2$ antennas at transmitters $1,2$ and $N_1, N_2$ antennas at receivers $1, 2$, respectively) and arbitrary partial CSIT levels. Specifically, we ask for the DoF that can be achieved by one user while the other user achieves his maximum possible interference-free DoF. The focus here is on achievable schemes, leaving the outer bounds for future work. As one might expect, signal space partitioning becomes a much more sophisticated in a MIMO setting. As a highlight, we note the need for ``elevated multiplexing", i.e., spreading of signals across transmit antennas at elevated power levels.  A remarkable consequence of elevated multiplexing is that there is a DoF benefit from increasing the number of antennas at a transmitter even when it already has more antennas than its desired receiver and no CSIT is available to the transmitter. 

Finally, a convergence of research interests in this topic is evident. We note that within one week of the submission of this work to IEEE SPAWC on March 18, 2016, another work was posted on ArXiv on March 24,  (revised March 25) by Hao, Rasouli and Clerckx \cite{Hao_Rassouli_Clerckx} which independently studies the DoF achieved in the 2 user MIMO IC under partial CSIT. Hao et al. have a broader focus in \cite{Hao_Rassouli_Clerckx} (includes both MIMO IC and MIMO BC, DoF region) than what we pursue in this work (MIMO IC, corner points of DoF region where one user achieves his maximum DoF). It is an interesting exercise to directly compare the two results where they overlap. Remarkably, a direct comparison shows that our signal space partitioning approach is in general strictly stronger than the approach taken by \cite{Hao_Rassouli_Clerckx}. In particular, the difference is evident in the class of channels where $M_1<N_1\leq N_2<M_2$. From this broad class of channels, a representative example is included in Fig. \ref{fig:ex4}, where $(M_1, M_2, N_1, N_2)=(1,4,2,3)$. 

%

{\it Notation:} We define $(A)^+=\max(A, 0)$. $\min^+(A, B)$ is defined as follows
$$
{\min}^+(A, B)=\begin{cases}
\min(A, B), &\mbox{if}\ \min(A, B)\ge0\\
0, &\mbox{if}\ \min(A, B)<0
\end{cases}
$$

\section{System Model}\label{sec_model}
Consider a 2-user Gaussian MIMO interference channel, where transmitters 1, 2 are equipped with $M_1$, $M_2$ antennas, respectively, and receivers 1, 2 are equipped with $N_1$, $N_2$ antennas, respectively.  Each transmitter wishes to send an independent message to its corresponding receiver.  At time slot $t\in\mathbb{N}$, the channel input-output equations are given by 
\begin{align}
{Y}_{1}(t)=\mathbf{H}_{11}(t){X}_{1}(t)+\mathbf{H}_{12}(t){X}_{2}(t)+{Z}_{1}(t), \label{eq::received1}\\
{Y}_{2}(t)=\mathbf{H}_{21}(t){X}_{1}(t)+\mathbf{H}_{22}(t){X}_{2}(t)+{Z}_{2}(t),\label{eq::received2}
\end{align}
Here, ${X}_{k}(t)$ is the ${M_{k}\times1}$ signal vector sent from Transmitter $k$, $k\in\{1, 2\}$, which is subject to the power constraint $P$. ${Y}_{k}(t)$ is the ${N_{k}\times1}$ the received signal vector at Receiver $k$. ${Z}_{k}(t)$ is the ${N_{k}\times1}$ i.i.d. additive white Gaussian noise (AWGN) vector at Receiver $k$, each entry of which is an i.i.d. Gaussian random variable with zero-mean and unit-variance.
$\mathbf{H}_{ji}(t)$ is the ${N_{j}\times M_{i}}$ channel matrix from Transmitter $i$ to Receiver $j$. Under partial CSIT, channel matrices $\mathbf{H}_{ji}(t)$, $\forall i, j\in\{1, 2\}$, are represented as
\begin{equation}
\mathbf{H}_{ji}(t)=\hat{\mathbf{H}}_{ji}(t)+\sqrt{P^{-\beta_{ji}}}\tilde{\mathbf{H}}_{ji}(t)
\end{equation}
where $\hat{\mathbf{H}}_{ji}(t)$ is the ${N_{j}\times M_{i}}$ estimated channel matrix while $\tilde{\mathbf{H}}_{ji}(t)$ is the ${N_{j}\times M_{i}}$ estimation error matrix. We assume that the entries of $\hat{\mathbf{H}}_{ji}(t)$ and $\tilde{\mathbf{H}}_{ji}(t)$ are drawn from continuous joint distributions with bounded densities, with the difference that the actual realizations of $\hat{\mathbf{H}}_{ji}(t)$ are revealed to the transmitters, but the realizations of $\tilde{\mathbf{H}}_{ji}(t)$ are not available to the transmitter. To avoid degenerate conditions, the ranges of values of all channel coefficients are bounded away from infinity. The parameter $\beta_{ji}$ measures the quality of the  channel estimate. If $\beta_{ji}=0$, then it corresponds to the case when there is no current CSIT. If $\beta_{{ji}}\ge1$, then it corresponds to the case that the current CSIT is as good as perfect (for DoF). Throughout this paper, we assume that $\beta\in[0, 1]$.
 
Since codebooks, probability of error, achievable rates ($R_1$, $R_2$) and capacity region $\mathcal{C}(P)$ are all defined in the standard Shannon theoretic sense, their definitions will not be repeated here.  The DoF tuple ($d_1$, $d_2$) is said to be achievable if there exist $(R_1(P), R_2(P))\in\mathcal{C}(P)$ such that
\begin{align}
d_{1}=&\lim_{p \to \infty}\frac{R_1(P)}{\log(P)},\\
d_{2}=&\lim_{p \to \infty}\frac{R_2(P)}{\log(P)}.
\end{align}

We are interested in the DoF achievable by a user while the other user is achieving his interference-free maximum DoF. To this end, without loss of generality, we will assume, that User 1 achieves $d_1=\min(M_1,N_1)$ DoF, and explore the DoF that are simultaneously achievable by User $2$.

\section{Examples}
Before stating the general result, we present a few  examples that highlight key ideas,  in particular what we mean by ``elevated multiplexing". Consider a transmitter that has no CSIT, and has as many antennas as its desired receiver. Is there a DoF benefit from further increasing the number of antennas at such a transmitter? Additional antennas are typically useful for zero-forcing or interference alignment. Since the absence of CSIT makes both zero-forcing and interference alignment impossible for this transmitter, one might expect that additional transmit antennas  bring no DoF benefit. The following examples shows that indeed there is a DoF benefit from additional antennas, and the key to this counterintuitive outcome is the idea of elevated multiplexing.

\subsection{$(M_1,M_2,N_1,N_2)=(1,4,1,3)$}\label{example1} 
\begin{figure}[h]
\centerline{\includegraphics[width=0.5\textwidth]{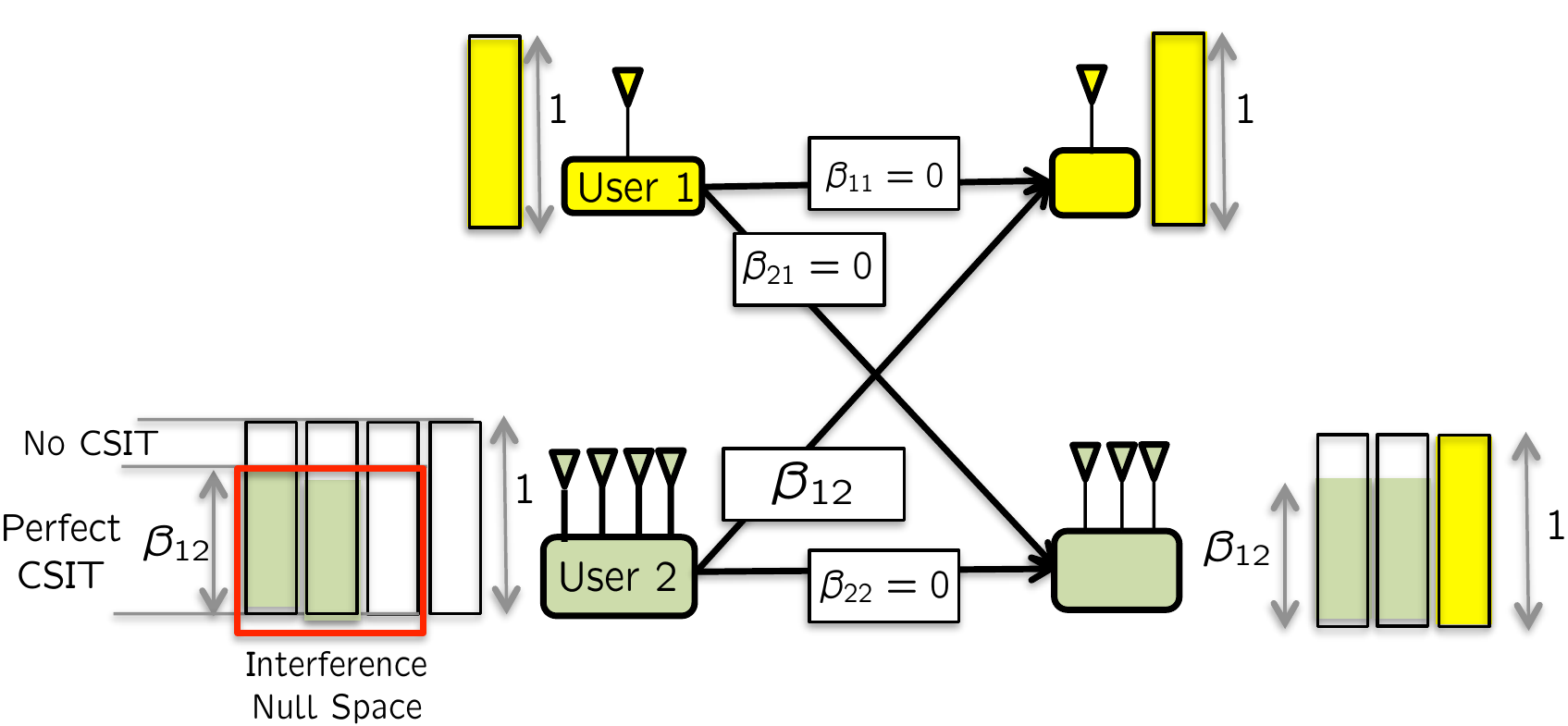}}
\caption{$(M_1,M_2,N_1,N_2)=(1,4,1,3)$, $(d_1,d_2)=(1,2\beta_{12})$.}
\end{figure}
Let us start with the setting $(M_1,M_2, N_1, N_2)=(1,4,1,3)$, where User $1$ achieves $d_1=1$, i.e., his maximum DoF. Suppose Transmitter $2$ has partial CSIT level $\beta_{12}$ for his interference carrying link to Receiver $1$, but  no other CSIT is available, i.e., all other $\beta_{ij}=0$. Since $d_1=1$, the signal from User $1$ occupies one full spatial dimension at both receivers. At Receiver $1$, this exhausts the desired signal space, so any interference from User $2$ should not rise above the noise floor (in the DoF sense). The only signal space this leaves for Transmitter $2$ is the null-space of the estimated channel $\hat{\bf H}_{12}$, within which Transmitter $2$ must not exceed the power level $P^{\beta_{12}}$. Only $2$ dimensions are left free from interference at Receiver $2$, and the desired signal power in each dimension is $P^{\beta_{12}}$. So User $2$ achieves $d_2=2\beta_{12}$.

Mathematically, the transmitted signals are,
\begin{eqnarray*}
X_1&=&\sqrt{P}\tilde{X}_1\\
X_2&=&\sqrt{P^{\beta_{12}}}(V_{21} \hat{X}_{21}+V_{22}\hat{X}_{22})
\end{eqnarray*}
Here $\tilde{X}_1, \hat{X}_{21}, \hat{X}_{22}$ are independent Gaussian codewords from unit power codebooks which carry $1, \beta_{12}, \beta_{12}$ DoF, respectively. $V_{21}, V_{22}$ are $4\times 1$ unit vectors in the null space of $\hat{\bf H}_{12}$, i.e., 
\begin{eqnarray*}
\hat{\bf H}_{12}[V_{21}~~~V_{22}]&=&[0~~0]
\end{eqnarray*}
The received signals are
\begin{eqnarray*}
Y_1&=&\sqrt{P}{\bf H}_{11}\tilde{X}_1+Z_1\nonumber\\
&&+\sqrt{P^{\beta_{12}}}(\hat{\bf H}_{12}+\sqrt{P^{-\beta_{12}}}\tilde{\bf H}_{12})(V_{21} \hat{X}_{21}+V_{22}\hat{X}_{22})\\
&=&\sqrt{P}{\bf H}_{11}\tilde{X}_1+O(1)+Z_1\\
Y_2&=&\sqrt{P}{\bf H}_{21}\tilde{X}_1+\sqrt{P^{\beta_{12}}}{\bf H}_{22}(V_{21} \hat{X}_{21}+V_{22}\hat{X}_{22})+Z_2
\end{eqnarray*}
Thus, $(d_1,d_2)=(1,2\beta_{12})$ is achieved. Incidentally, in this channel if $d_1=1$, then the maximum possible DoF for User 2 with no CSIT ($\beta_{12}=0$) is $\tilde{d}_2=0$, and with perfect CSIT  ($\beta_{12}=1$) is $\hat{d}_2=2$. Therefore, the subspace partition scheme presented above achieves $d_2=\beta_{12}\hat{d}_2+(1-\beta_{12})\tilde{d}_2$ DoF, which can be shown to be optimal.

\subsection{$(M_1,M_2,N_1,N_2)=(3,4,1,3)$}\label{example2} 
\begin{figure}[h]
\centerline{\includegraphics[width=0.5\textwidth]{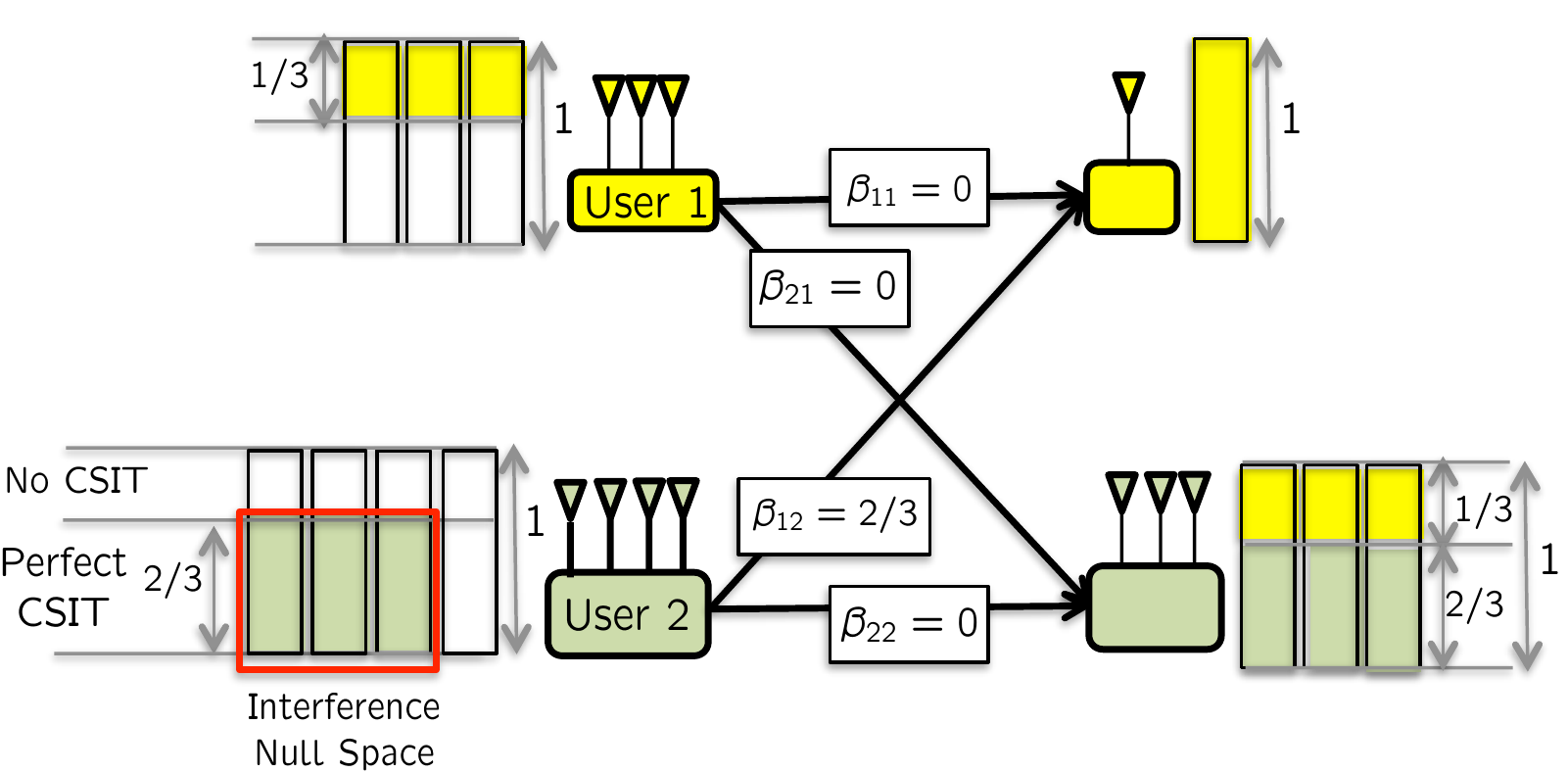}}
\caption{$(M_1,M_2,N_1,N_2)=(3,4,1,3)$. Elevated multiplexing at Transmitter $1$ helps achieve $(d_1,d_2)=(1,\min(2,3\beta_{12}))$.}\label{fig:ex1}
\end{figure}
Even though in the previous example Transmitter $1$ already has as many antennas as Receiver $1$,  let us further increase the number of transmit antennas to $M_1=3$, while keeping everything else the same, so Transmitter $1$ still has no CSIT and $d_1=1$.  To further simplify the exposition, let us consider specifically $\beta_{12}=2/3$. Remarkably, as shown in Fig. \ref{fig:ex1},  it is now possible for User $2$ to achieve $2$ DoF (same as with perfect CSIT). To accomplish this, User $1$ multiplexes his $1$ DoF into three streams, each carrying $1/3$ DoF and transmits them from its three antennas, each with elevated power level $\sim P$. At the same time, User $2$ transmits three streams, each with power $P^{2/3}$ along the three dimensions that are in the null space of his estimated channel to Receiver $1$. As before this signal space partitioning ensures that the interference caused at Receiver $1$ from Transmitter $2$ remains at the noise floor level. In the absence of interference, Receiver $1$ \emph{jointly} decodes the  three desired streams from Transmitter $1$ as a multiple access channel (MAC). Receiver $2$ first decodes the interfering signal from Transmitter $1$ by treating its own desired signals as noise. Each of the three desired streams is received at power level $\sim P^{2/3}$ while each of the undesired streams is received at power level $P$, so the SINR for each stream is $P/P^{2/3}=P^{1/3}$. Since each interfering stream carries only $1/3$ DoF, and Receiver $2$ has $3$ antennas to separate the streams, it is able to decode and subsequently remove all interference. This leaves only the desired signal streams, which are then decoded to achieve $d_2=2/3\times 3=2$ DoF for User $2$. Note that this is clearly optimal, in fact it is also the best possible DoF for User $2$ even if perfect CSIT was available to both transmitters. Also note the role of elevated multiplexing at Transmitter $1$ which has no CSIT. Because of this elevated multiplexing, Receiver $1$ is able to resolve the three streams jointly in its one interference-free received dimension, while Receiver $2$ is able to simultaneously resolve the three streams separately in its $3$ received dimensions, each of which sees an elevated noise floor (due to his desired signals) of $P^{2/3}$.

Generalization to other values of $\beta_{12}$ is straightforward. If $\beta_{12}>2/3$ then $(d_1,d_2)=(1,2)$ is still trivially achievable because improved CSIT cannot hurt. If $\beta_{12}<2/3$ then Transmitter $2$ sets the power level and the DoF of each of his $3$ streams as $\beta_{12}$ to keep the interference at Receiver $1$ below the noise floor. The signals are decoded as before at each receiver, to achieve the DoF tuple $(d_1,d_2)=(1,3\beta_{12})$.

\subsection{$(M_1,M_2,N_1,N_2)=(2,4,1,3)$. }\label{example3} 
\begin{figure}[h]
\centerline{\includegraphics[width=0.5\textwidth]{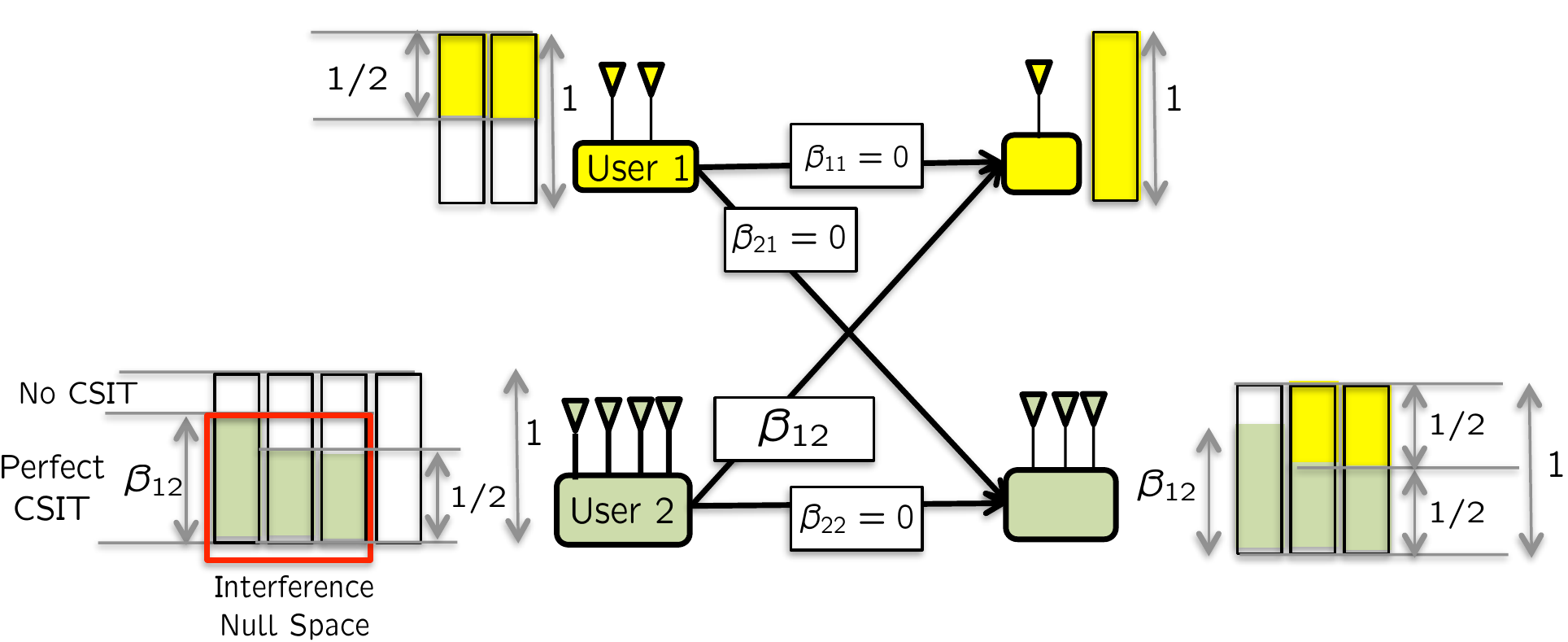}}
\caption{$(M_1,M_2,N_1,N_2)=(2,4,1,3)$. Elevated multiplexing at Transmitter $1$ helps achieve $(d_1,d_2)=(1,\min(1+\beta_{12},3\beta_{12}))$.}\label{fig:ex3}
\end{figure}
Consider now that Transmitter $1$ has $M_1=2$ antennas, while all other assumptions remain the same. In this case, if User $1$ achieves $d_1=1$ DoF, User $2$ can simultaneously achieve $d_2=\min(1+\beta_{12},3\beta_{12})$ DoF as shown in Fig. \ref{fig:ex3}.  To accomplish this, User $1$ multiplexes his $1$ DoF into $2$ streams, each carrying $1/2$ DoF and transmits them from its two antennas, each with elevated power level $\sim P$. At the same time, User $2$ transmits three streams, the first with power $P^{\beta_{12}}$ and the next two with power $P^{\min(1/2,\beta_{12})}$, along the three dimensions that are in the null space of his estimated channel to Receiver $1$ ($\beta_{12}>1/2$ in Fig. \ref{fig:ex3}). As before this signal space partitioning ensures that the interference caused at Receiver $1$ from Transmitter $2$ remains at the noise floor level. In the absence of interference, Receiver $1$ \emph{jointly} decodes the  two desired streams from Transmitter $1$ as a multiple access channel (MAC). Receiver $2$ first decodes the interfering signal from Transmitter $1$ in the two dimensional space orthogonal to its own first desired stream, by treating its remaining desired signals as noise. Each of the two remaining desired streams is received at power level $\sim P^{\min(1/2,\beta_{12})}$ while each of the undesired streams is received at power level $P$, so the SINR for each stream is $P^{1-\min(1/2,\beta_{12})}\geq P^{1/2}$. Since each interfering stream carries only $1/2$ DoF, Receiver $2$ is able to decode and subsequently remove all interference. This leaves only the desired signal streams, which are then decoded to achieve $d_2=\beta_{12}+\min(1/2,\beta_{12})+\min(1/2,\beta_{12})$ DoF for User $2$. 

\begin{figure}[h]
\centerline{\includegraphics[width=0.5\textwidth]{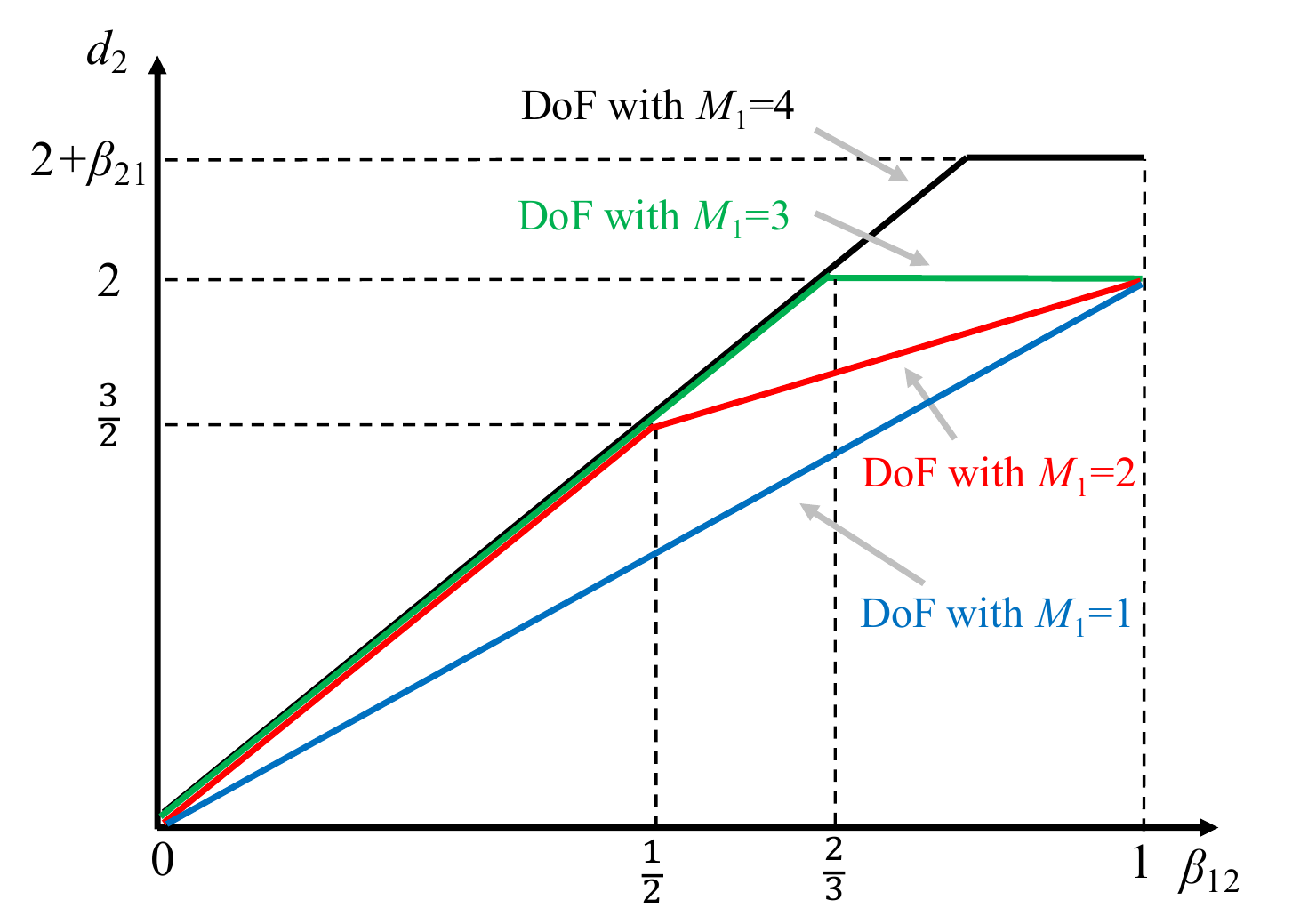}}
\caption{$(M_2,N_1,N_2)=(4,1,3)$. DoF achieved by User $2$ when User $1$ achieves his maximum DoF, $d_1=1$. Note that the DoF improve as $M_1$ increases even though $M_1\geq N_1$ and Transmitter $1$ has no CSIT.}\label{fig:dof1}
\end{figure}

The three examples discussed so far are summarized in Fig. \ref{fig:dof1}.

\subsection{$(M_1,M_2,N_1,N_2)=(1,4,2,3)$. }\label{example4} 
\begin{figure}[h]
\centerline{\includegraphics[width=0.5\textwidth]{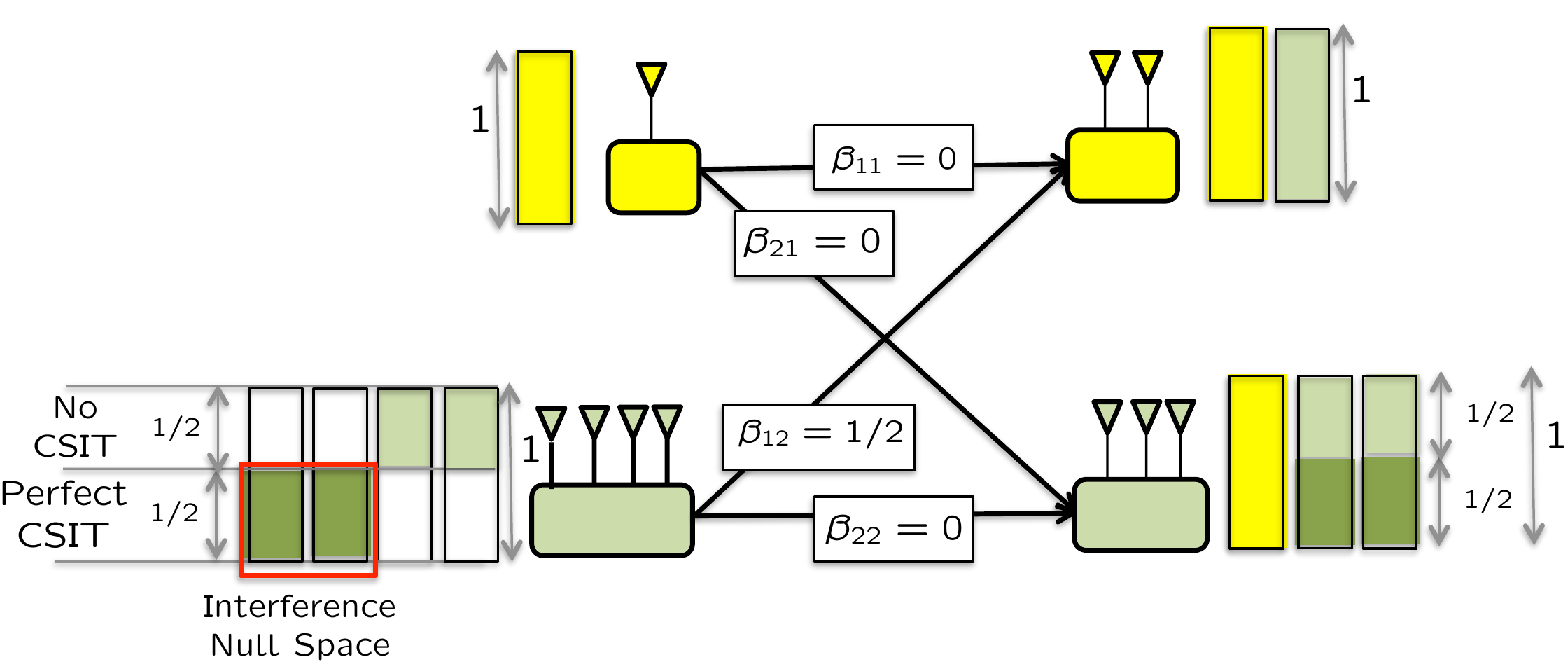}}
\caption{$(M_1,M_2,N_1,N_2)=(1,4,2,3)$. Elevated multiplexing at Transmitter $2$ helps achieve $(d_1,d_2)=(1,2)$.}\label{fig:ex4}
\end{figure}
\begin{figure}
\centering 
\includegraphics[width=3.5in]{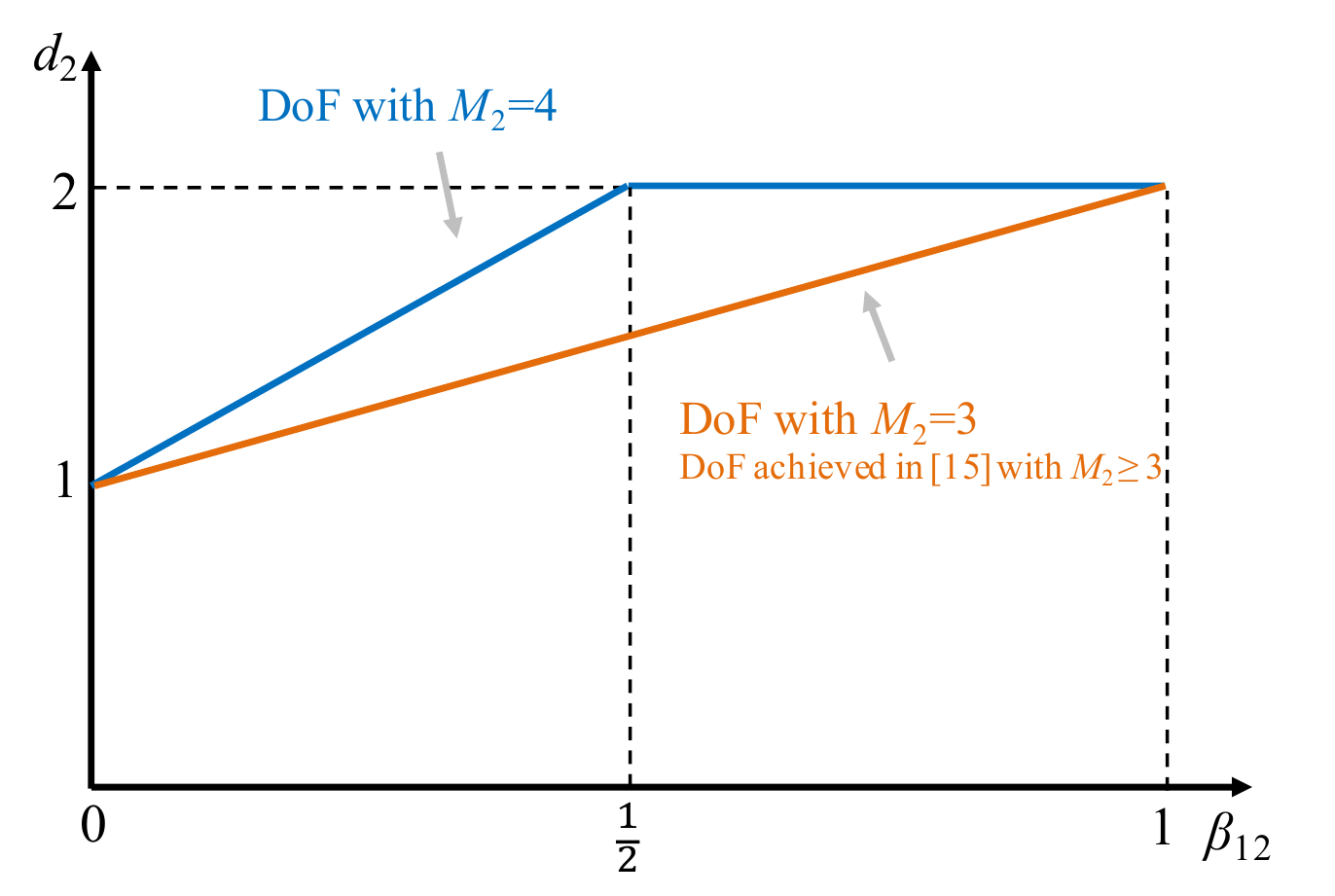}
\caption{The achievable DoF for User 2 as a function of $\beta_{12}$, when User 1 achieves its maximum DoF with $(M_1, N_1, N_2)=(1, 2, 3)$.}\label{fig:dof2}
\end{figure}

For the next example, we consider the setting $(M_1,M_2,N_1,N_2)=(1,4,2,3)$ where User $1$ has more receive antennas than transmit antennas. We consider $\beta_{12}=1/2$. Here $(d_1,d_2)=(1,2)$ is achieved as shown in Fig. \ref{fig:ex4}. This example shows how elevated multiplexing is useful at User $2$, in a way that the multiplexed signals are decoded separately in space by the desired receiver and jointly in signal levels by the undesired receiver. User $1$ simply sends his $1$ DoF carrying stream from his single transmit antenna at power level $\sim P$. Transmitter $2$ fully occupies the two dimensions in the null space of $\hat{\bf H}_{12}$, along which it can send at power levels up to $P^{1/2}$ without exceeding the noise floor at Receiver $1$. Since Receiver $1$ also has an extra dimension, Transmitter $2$ uses elevated multiplexing to send two more streams, each carrying $1/2$ DoF at elevated power levels of $\sim P$, along generic directions. At Receiver $1$, first the desired signal is zero forced and  the two elevated interference streams are jointly decoded in the remaining dimension. After these interfering streams are removed, the Receiver is able to decode its desired signal to recover the desired $d_1=1$ DoF. Receiver $2$ on the other hand, first zero forces the interference and in the remaining $2$ dimensions, first decodes the elevated streams while treating the other streams as noise. Since the elevated streams have power $\sim P$ while the other 2 desired streams have power $\sim P^{1/2}$, the SINR for this decoding is $P^{1/2}$ per dimension, which allows Receiver $2$ to decode and subtract both elevated streams. The remaining streams are then decoded separately along the two interference-free dimensions. The scheme is easily generalized to arbitrary $\beta_{12}$ values. The results for this example (and the case $M_2=3$ for comparison) are shown in Fig. \ref{fig:dof2}.
\subsection{$(M_1,M_2,N_1,N_2)=(4,4,1,3)$. }\label{example5} 
\begin{figure}[h]
\centerline{\includegraphics[width=0.5\textwidth]{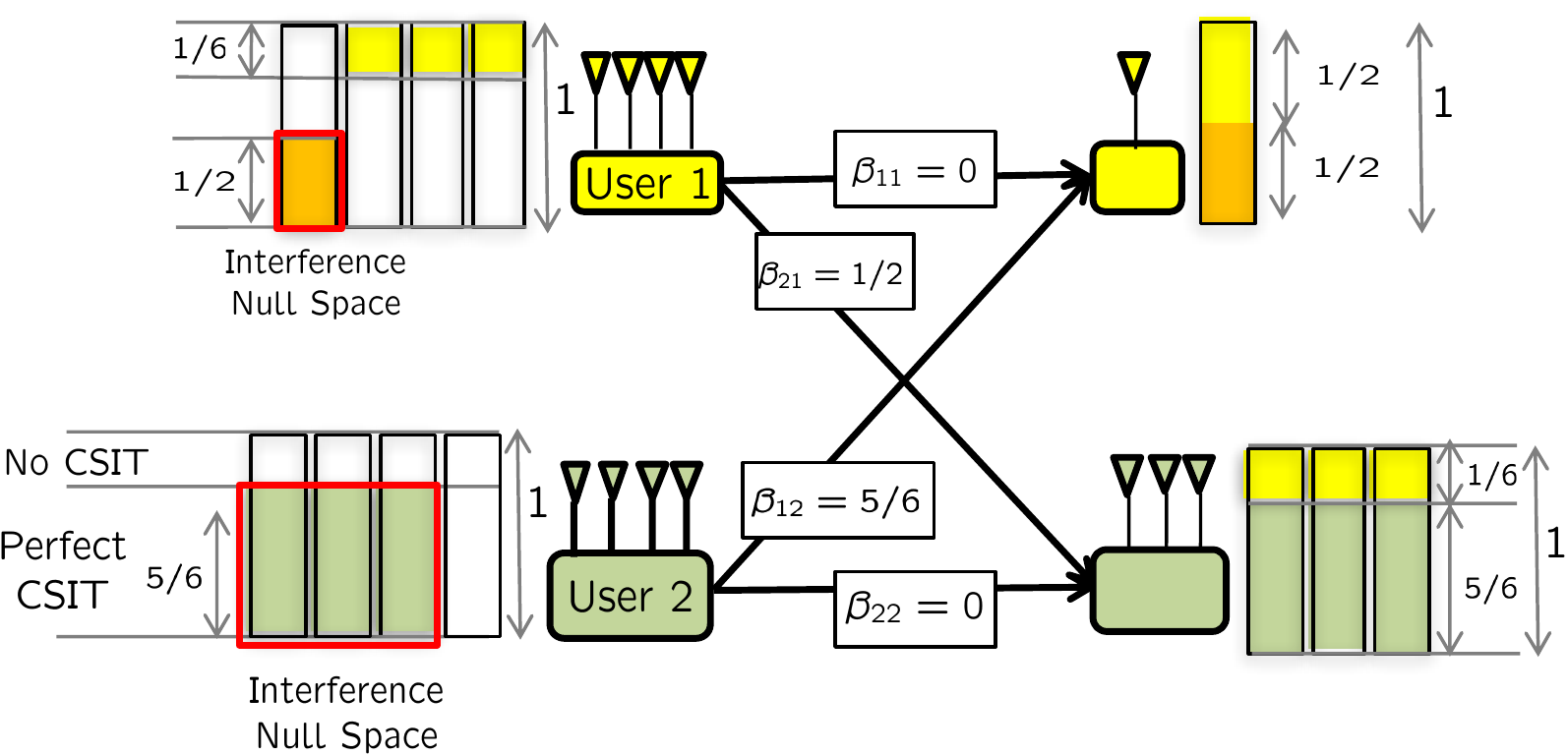}}
\caption{$(M_1,M_2,N_1,N_2)=(4,4,1,3)$. Elevated multiplexing at Transmitter $2$ helps achieve $(d_1,d_2)=(1,2.5)$.}\label{fig:ex5}
\end{figure}
While it turns out that the CSIT of desired channels is  irrelevant for  DoF, in general the DoF may depend on the partial CSIT level at \emph{both} cross channels, i.e., $\beta_{12}$ and $\beta_{21}$. Our final example, illustrates such a setting. Specifically we consider $(M_1,M_2,N_1,N_2)=(4,4,1,3)$  with $\beta_{12}=5/6$ and $\beta_{21}=1/2$, where $(d_1,d_2)=(1,5/2)$ is achieved as shown in Fig. \ref{fig:ex5}. Transmitters $1$ and $2$ fully exploit the null-space of the estimated channel to the undesired receiver, up to a power levels $P^{\beta_{21}}$ and $P^{\beta_{12}}$, respectively, which keeps this interference below the noise floor at the undesired receiver. Transmitter $1$ uses elevated multiplexing to obtain the remaining $1-\beta_{21}=1/2$ of its DoF by splitting into three streams which carry $1/6$ DoF each. At Receiver $1$ the multiplexed streams are jointly decoded to achieve $d_1=1$ DoF. At Receiver $2$ the interfering multiplexed streams are separated in spatial dimensions and decoded while treating its own desired signal as noise. Removing the decoded interference then allows Receiver $2$ to decode its desired signal to achieve $d_2=5/2$ DoF.  For arbitrary values of $\beta_{12}, \beta_{21}$, the scheme generalizes to achieve $(d_1,d_2)=(1,\min(3\beta_{12},2+\beta_{21}))$. 

\section{MIMO IC: General setting}\label{sec_result}
With the key ideas highlighted in the previous section through various examples, we are now ready to consider the MIMO interference channel with arbitrary number of antennas at each node and arbitrary levels of partial CSIT. The achievable DoF for this general setting are stated in the following theorem.

\begin{theorem}\label{theorem:dof}
{\it For the 2-user MIMO interference channel with partial CSIT, if User 1 achieves its interference-free DoF, i.e., $d_1=\min\{M_1, N_1\}$, the DoF value achieved by User 2 is
\begin{equation}
d_2={\min}^+\{A, B, C\}, \label{main}
\end{equation}
where
\begin{align}
A=&\min[\max(M_1, N_2), \max(M_2, N_1), M_1{+}M_2, N_1{+}N_2]\nonumber\\
&-\min(M_1, N_1),\\
B=&{\min}^+(N_1{-}M_1, M_2)+\min(M_1, N_2)-\min(M_1, N_1)\nonumber\\
&+\beta_{12}{\min}^+(N_2{-}M_1, M_2)+\beta_{21}{\min}^+(M_1{-}N_2,N_1),\\
C=&{\min}^+(N_1{-}M_1, M_2)+\beta_{12}{\min}^+(M_2{-}N_1, N_2).
\end{align}}
\end{theorem}
Note that in a same channel, if $d_2=\min(M_2, N_2)$, then the DoF value achieved by User 1 can be obtained by just switching the indices in Theorem 1.

Let us conclude with some high level insights into the theorem. First,  we note that in the theorem, $A$ corresponds to the DoF achieved by User $2$ with perfect CSIT,  $B$ corresponds to the restrictions at Receiver $2$ needed to decode all the desired messages. $C$ corresponds to the maximum DoF that can be sent by Transmitter $2$ without hurting User $1$. 

Note also that the result matches the known DoF for two extreme cases, i.e., MIMO IC with perfect CSIT \cite{Jafar_Fakhereddin} when all $\beta_{ji}=1$ and MIMO IC with no CSIT \cite{Zhu_Guo_MIMOIC} when all $\beta_{ji}=0$. However, more significantly, the result is not a simple extension of the two extreme cases.  

Next we note that the DoF do not depend on $\beta_{11}, \beta_{22}$, i.e., the channel knowledge of desired links is not critical. This observation is consistent with the understanding of interference alignment and zero forcing schemes based on all previous studies. 

Another remarkable observation is how the CSIT requirement changes with the null space of cross-links. It is clear that if there is no null space for the channel from Transmitter $i$ to Receiver $j$, i.e., $M_i\ge N_j$, then the achievable DoF do not depend on $\beta_{ji}$, i.e., CSIT for this cross-link is not needed.

\subsection{Proof for Achievability} 
We will consider  the case where $d_1=\min(M_1,N_1)$ and also the case where $d_2=\min(M_2,N_2)$. In each case we will determine the achievable DoF of the other user. With this approach we can assume with no loss of generality  that   $N_1\le N_2$. Then the parameter space of Theorem 1 can be  divided into the following four cases. 
\subsubsection{Case 1: $M_2\le N_2$}
$M_2\le N_1$ is trivial because the DoF is the same as that with both perfect and no CSIT. Therefore, let us consider $M_2>N_1$, so that \eqref{main} becomes
\begin{equation}
d_2=(N_1{-}M_1)^++\beta_{12}(M_2-N_1).\label{case1}
\end{equation}
With $d_1=\min(M_1, N_1)$, \eqref{case1} can be achieved with only partial zero-forcing precoding. In each channel use, User 1  sends  $\min(M_1, N_1)$  streams, each carrying 1 DoF, and each with power level $\sim P$. Transmitter 2 fully occupies the $M_2-N_1$ dimensions in the null space of $\hat{\mathbf{H}}_{12}$, along which it can send at power levels up to $P^{\beta_{12}}$ without exceeding the noise floor at Receiver 1. Since Receiver 1 also has $(N_1{-}M_1)^+$ extra dimensions, Transmitter 2 sends $(N_1{-}M_1)^+$ additional streams, each carrying 1 DoF at power levels of  $P$, along generic directions.

Mathematically, the transmitted signals are,
\begin{align}
X_1=&c_o\sqrt{P}\sum_{l=1}^{\min(M_1, N_1)} V^{c}_{1l}X^{c}_{1l}\\
X_2=&c_1\sqrt{P}\sum_{i=1}^{(N_1{-}M_1)^+} V^{c}_{2i}X^{c}_{2i}+c_2\sqrt{P^{\beta_{12}}}\sum_{j=1}^{M_2{-}N_1} V^{p}_{2j}X^{p}_{2j}
\end{align}
Here $X^{c}_{11}$, $X^{c}_{12}$, $\cdots$, $X^{c}_{1\min(M_1, N_1)}$ and $X^{c}_{21}$, $\cdots$, $X^{c}_{2(N_1{-}M_1)^+}$ are independent Gaussian codewords from unit power codebooks, each carries 1 DoF, and the superscript `c' is used to indicate that these codewords can be decoded by both receivers (common). $X^{p}_{21}$, $\cdots$, $X^{p}_{2(M_2{-}N_1)}$ are independent Gaussian codewords from unit power codebooks, each carries $\beta_{12}$ DoF, and the superscript $p$ is used to indicate that these are `private', i.e., only decoded by the intended receiver, in this case User 2.
$c_o$, $c_1$ and $c_2$ are scaling factors, $O(1)$ in $P$, chosen to ensure that the transmit power constraint is satisfied. $V^{c}_{1l}$ and $V^{c}_{2i}$ are $M_1\times1$ and $M_2\times1$ generic unit vectors, respectively. $V^{p}_{2j}$ are $M_2\times1$ unit vectors chosen so that
\begin{align}
\hat{\mathbf{H}}_{12}\begin{bmatrix}
V^{p}_{21}  & V^{p}_{22}   & \cdots & V^{p}_{2(M_2{-}N_1)}  
\end{bmatrix}={\bf 0}
\end{align}

The received signals are 
\begin{align}
Y_1=&c_o\sqrt{P}{\bf H}_{11}\sum_{l=1}^{\min(M_1, N_1)} V^{c}_{1l}X^{c}_{1l}+c_1\sqrt{P}{\bf H}_{12}\sum_{i=1}^{(N_1{-}M_1)^+} V^{c}_{2i}X^{c}_{2i}\nonumber\\
&+c_2\sqrt{P^{\beta_{12}}}(\hat{\bf H}_{12}+\sqrt{P^{-\beta_{12}}}\tilde{\bf H}_{12})\sum_{j=1}^{M_2{-}N_1} V^{p}_{2j}X^{p}_{2j}+Z_1\\
=&c_o\sqrt{P}{\bf H}_{11}\sum_{l=1}^{\min(M_1, N_1)} V^{c}_{1l}X^{c}_{1l}+c_1\sqrt{P}{\bf H}_{12}\sum_{i=1}^{(N_1{-}M_1)^+} V^{c}_{2i}X^{c}_{2i}+O(1)+Z_1\\
Y_2=&c_o\sqrt{P}{\bf H}_{21}\sum_{l=1}^{\min(M_1, N_1)} V^{c}_{1l}X^{c}_{1l}+c_1\sqrt{P}{\bf H}_{22}\sum_{i=1}^{(N_1{-}M_1)^+} V^{c}_{2i}X^{c}_{2i}+c_2\sqrt{P^{\beta_{12}}}{\bf H}_{22}\sum_{j=1}^{M_2{-}N_1} V^{p}_{2j}X^{p}_{2j}+Z_2
\end{align}

Since $\min(M_1, N_1)+(N_1{-}M_1)^+=N_1$, Receiver 1 has enough antennas to decode all the streams carrying common messages by treating other signals as white noise. Similarly, Receiver 2 has enough antennas to decode all the streams separately due to $\min(M_1, N_1)+(N_1{-}M_1)^++M_2{-}N_1\le N_2$.

{\color{black}On the other hand, if $d_2=\min(M_2, N_2)=M_2$, then $d_1=0$ is  trivially achieved.}

\subsubsection{Case 2: $M_1{<}N_1{\le} N_2{<}M_2$} 
In this case if $d_1=M_1$ then \eqref{main} becomes
\begin{equation}
d_2{=}\min[N_2-M_1, N_1-M_1+\beta_{12}\min(M_2-N_1, N_2-M_1)].\label{case2}
\end{equation}
The examples in Section \ref{example4} correspond to this case. To achieve \eqref{case2}, not only partial zero-forcing precoding, but also the elevated multiplexing is required at Transmitter 2.
In each channel use, User 1 simply sends  $M_1$  streams from his $M_1$ transmit antennas, each carrying $1$ DoF, and each at power level $\sim P$. Transmitter 2 occupies $\min(M_2-N_1, N_2-M_1)$ dimensions in the null space of $\hat{\mathbf{H}}_{12}$, along which it can send at power levels up to $P^{\bar{\beta}_{12}}$ without exceeding the noise floor at Receiver 1, where $\bar{\beta}_{12}$ is define as $\bar{\beta}_{12}=\min(\beta_{12}, \tfrac{N_2-N_1}{\min(M_2-N_1, N_2-M_1)})$. Since Receiver 1 also has $N_1{-}M_1$ extra dimensions, Transmitter 2 uses elevated multiplexing to send $M_2$ more streams, each carrying $\tfrac{N_1-M_1}{M_2}$ DoF at power levels of  $P$, along generic directions.

Mathematically, the transmitted signals are,
\begin{align}
X_1=&c_o\sqrt{P}\sum_{l=1}^{M_1} V^{c}_{1l}X^{c}_{1l}\\
X_2=&c_1\sqrt{P}\sum_{i=1}^{M_2} V^{c}_{2i}X^{c}_{2i}+c_2\sqrt{P^{\bar{\beta}_{12}}}\sum_{j=1}^{\min(M_2-N_1, N_2-M_1)} V^{p}_{2j}X^{p}_{2j}
\end{align}
Here $X^{c}_{11}$, $X^{c}_{12}$, $\cdots$, $X^{c}_{1M_1}$ and $X^{c}_{21}$, $\cdots$, $X^{c}_{2M_2}$ are independent Gaussian codewords from unit power codebooks that can be decoded by both receivers. Each $X^{c}_{1l}$ carries 1 DoF while each $X^{c}_{2i}$ carries $\tfrac{N_1-M_1}{M_2}$ DoF. $X^{p}_{21}$, $\cdots$, $X^{p}_{2\min(M_2-N_1, N_2-M_1)}$ are independent Gaussian codewords from unit power codebooks, each carries $\bar{\beta}_{12}$ DoF that to be decoded only by User 2.
$c_o$, $c_1$ and $c_2$ are scaling factors, $O(1)$ in $P$, chosen to ensure that the transmit power constraint is satisfied.

Here $V^{c}_{1l}$ and $V^{c}_{2i}$ are $M_1\times1$ and $M_2\times1$ generic unit vectors, respectively. $V^{p}_{2j}$ is a $M_2\times1$ unit vector chosen so that
\begin{align}
\hat{\mathbf{H}}_{12}\begin{bmatrix}
V^{p}_{21}  & V^{p}_{22}   & \cdots & V^{p}_{2\min(M_2-N_1, N_2-M_1)}  
\end{bmatrix}={\bf 0}
\end{align}

The received signals are 
\begin{align}
Y_1=&c_o\sqrt{P}{\bf H}_{11}\sum_{l=1}^{M_1} V^{c}_{1l}X^{c}_{1l}+c_1\sqrt{P}{\bf H}_{12}\sum_{i=1}^{M_2} V^{c}_{2i}X^{c}_{2i}\nonumber\\
&+c_2\sqrt{P^{\bar{\beta}_{12}}}(\hat{\bf H}_{12}+\sqrt{P^{-\beta_{12}}}\tilde{\bf H}_{12})\sum_{j=1}^{\min(M_2-N_1, N_2-M_1)} V^{p}_{2j}X^{p}_{2j}+Z_1\\
=&c_o\sqrt{P}{\bf H}_{11}\sum_{l=1}^{M_1} V^{c}_{1l}X^{c}_{1l}+c_1\sqrt{P}{\bf H}_{12}\sum_{i=1}^{M_2} V^{c}_{2i}X^{c}_{2i}+O(1)+Z_1\\
Y_2=&c_o\sqrt{P}{\bf H}_{21}\sum_{l=1}^{M_1} V^{c}_{1l}X^{c}_{1l}+c_1\sqrt{P}{\bf H}_{22}\sum_{i=1}^{M_2} V^{c}_{2i}X^{c}_{2i}+c_2\sqrt{P^{\bar{\beta}_{12}}}{\bf H}_{22}\sum_{j=1}^{\min(M_2-N_1, N_2-M_1)} V^{p}_{2j}X^{p}_{2j}+Z_2
\end{align}

At Receiver 1, first the signals from Transmitter 1 are zero forced and the $M_2$ elevated streams are jointly decoded in the remaining $N_1-M_1$ dimensions as a MAC channel. After all the $X^{c}_{2i}$ are removed, Receiver 1 is able to decode the $M_1$ signals from Transmitter 1, i.e., $X^{c}_{1l}$. Receiver 2 on the other hand, first zero forced the signals from Transmitter 1 and in the remaining $N_2-M_1$ dimensions, first jointly decode $M_2$ elevated streams as a MAC channel while treating other $\min(M_2-N_1, N_2-M_1)$ streams carrying $X^{p}_{2j}$ as noise. After $X^{c}_{2i}$ are decoded and removed, $X^{p}_{2j}$ are then decoded separately along $\min(M_2-N_1, N_2-M_1)$ interference-free dimensions. 


On the other hand, in this setting, if $d_2=N_2$ then $d_1=0$ can be trivially achieved.

\subsubsection{Case 3: $N_1{\le} M_1{\le} N_2{<}M_2$} 
In this case if $d_1=N_1$ then \eqref{main} becomes
\begin{equation}
d_2{=}\min[\beta_{12}\min(M_2-N_1, N_2), M_1-N_1+\beta_{12}(N_2-M_1)].\label{case3}
\end{equation}
The examples in Section \ref{example1}, \ref{example2} and \ref{example3} correspond to this case. Partial zero-forcing precoding at User 2 is required. What's more, to help User 2 to achieve \eqref{case3}, User 1 needs to use elevated multiplexing. Specifically, Transmitter 1 multiplexes his $N_1$ DoF into $M_1$ streams, each carrying $\tfrac{N_1}{M_1}$ DoF with elevated power level $\sim P$. At the same time, User 2 occupies $\min(M_2-N_1, N_2)$ dimensions in the null space of $\hat{\mathbf{H}}_{12}$, the first $N_2-M_1$ streams are sent with power levels up to $P^{\beta_{12}}$ and the rest $M_1-N_2+\min(M_2-N_1, N_2)$ streams are sent with power levels up to $P^{\bar{\beta}_{12}}$ without exceeding the noise floor at Receiver 1, where $\bar{\beta}_{12}$ is defined as $\bar{\beta}_{12}=\min(\beta_{12}, \tfrac{M_1-N_1}{M_1-N_2+\min(M_2-N_1, N_2)})$.

Mathematically, the transmitted signals are,
\begin{align}
X_1=&c_o\sqrt{P}\sum_{l=1}^{M_1} V^{c}_{1l}X^{c}_{1l}\\
X_2=&c_1\sqrt{P^{\beta_{12}}}\sum_{j=1}^{N_2-M_1} V^{p}_{2j}X^{p}_{2j}+c_2\sqrt{P^{\bar{\beta}_{12}}}\sum_{i=N_2-M_1+1}^{\min(M_2-N_1, N_2)} V^{p}_{2i}X^{p}_{2i}
\end{align}
Here $X^{c}_{11}$, $X^{c}_{12}$, $\cdots$, $X^{c}_{1M_1}$ are independent Gaussian codewords from unit power codebooks that can be decoded by both receivers. Each $X^{c}_{1l}$ carries $\tfrac{N_1}{M_1}$ DoF. $X^{p}_{21}$, $\cdots$, $X^{p}_{2\min(M_2-N_1, N_2)}$ are independent Gaussian codewords from unit power codebooks that are to be decoded only by User 2. Each $X^{p}_{2j}$ and $X^{p}_{2i}$ carries $\beta_{12}$ and $\bar{\beta}_{12}$ DoF, respectively.
$c_o$, $c_1$ and $c_2$ are scaling factors, $O(1)$ in $P$, chosen to ensure that the transmit power constraint is satisfied.

Here $V^{c}_{1l}$ are $M_1\times1$ generic unit vectors. $V^{p}_{2j}$ and $V^{p}_{2i}$ are $M_2\times1$ unit vectors chosen so that
\begin{align}
\hat{\mathbf{H}}_{12}\begin{bmatrix}
V^{p}_{21}  & V^{p}_{22}   & \cdots & V^{p}_{2\min(M_2-N_1, N_2)}  
\end{bmatrix}={\bf 0}
\end{align}

The received signals are 
\begin{align}
Y_1=&c_o\sqrt{P}{\bf H}_{11}\sum_{l=1}^{M_1} V^{c}_{1l}X^{c}_{1l}+(\hat{\bf H}_{12}+\sqrt{P^{-\beta_{12}}}\tilde{\bf H}_{12})X_2+Z_1\\
=&c_o\sqrt{P}{\bf H}_{11}\sum_{l=1}^{M_1} V^{c}_{1l}X^{c}_{1l}+O(1)+Z_1\\
Y_2=&c_o\sqrt{P}{\bf H}_{21}\sum_{l=1}^{M_1} V^{c}_{1l}X^{c}_{1l}+c_1\sqrt{P^{\beta_{12}}}{\bf H}_{22}\sum_{j=1}^{N_2-M_1} V^{p}_{2j}X^{p}_{2j}+c_2\sqrt{P^{\bar{\beta}_{12}}}{\bf H}_{22}\sum_{i=N_2-M_1+1}^{\min(M_2-N_1, N_2)} V^{p}_{2i}X^{p}_{2i}+Z_2
\end{align}

As before the signal space partitioning ensures that the interference caused at Receiver 1 from Transmitter 2 remains at the noise floor level. In the absence of interference, Receiver 1 can jointly decode the $M_1$ desired streams from Transmitter 1 as a MAC channel. 

At the same time, Receiver 2 zero forces the first $N_2-M_1$ signals from Transmitter 2, i.e., $X^{p}_{2j}$. In the remaining $M_1$ dimensions, $M_1$ elevated streams from Transmitter 1 can be decoded (see Lemma \ref{lemma:mac} in the Appendix) by treating the rest $M_1-N_2+\min(M_2-N_1, N_2)$ streams carrying $X^{p}_{2i}$ as white noise. After $X^{c}_{1l}$ are decoded and removed, all the remaining signals can then be decoded separately along $\min(M_2-N_1, N_2)$ interference-free dimensions.

On the other hand, in this setting, if $d_2=N_2$ then $d_1=0$ can be trivially achieved.

\subsubsection{Case 4: $N_1{\le}N_2{<}\min(M_1, M_2)$} 
In this case if $d_1=N_1$ then \eqref{main} becomes
\begin{align}
d_2=\min[&\beta_{12}\min(M_2{-}N_1, N_2), N_2-N_1+\beta_{21}\min(M_1{-}N_2, N_1)].\label{case4}
\end{align}
This case can be seen as an extension of Case 3 where there is null space for the channel from Transmitter 1 to Receiver 2, The examples in Section \ref{example5} correspond to this case. Thus $d_2$ depends on both $\beta_{12}$ and $\beta_{21}$. To achieve \eqref{case4}, the only difference is that User 1 needs both partial zero-forcing precoding and elevated multiplexing to help User 2.

Specifically, Transmitter 1 occupies $\min(M_1-N_2, N_1)$ dimensions in the null space of $\hat{\mathbf{H}}_{21}$, along which it can send at power levels up to $P^{\beta_{21}}$ without exceeding the noise floor at Receiver 2. Then Transmitter 1 multiplexes his remaining $N_1-\beta_{21}\min(M_1-N_2, N_1)$ DoF into $M_1$ streams, each carrying $\tfrac{N_1-\beta_{21}\min(M_1-N_2, N_1)}{M_1}$ DoF with elevated power level $\sim P$. At the same time, Transmitter 2 occupies $\min(M_2-N_1, N_2)$ dimensions in the null space of $\hat{\mathbf{H}}_{12}$, along which it can send at power levels up to $P^{\bar{\beta}_{12}}$ without exceeding the noise floor at Receiver 2, where $\bar{\beta}_{12}$ is defined as $\bar{\beta}_{12}=\min(\beta_{12}, \tfrac{N_2-N_1+\beta_{21}\min(M_1-N_2, N_1)}{\min(M_2-N_1, N_2)})$.

Mathematically, the transmitted signals are,
\begin{align}
X_1=&c_o\sqrt{P}\sum_{l=1}^{M_1} V^{c}_{1l}X^{c}_{1l}+c_1\sqrt{P^{\beta_{21}}}\sum_{k=1}^{\min(M_1-N_2, N_1)} V^{p}_{1k}X^{p}_{1k}\\
X_2=&c_2\sqrt{P^{\bar{\beta}_{12}}}\sum_{i=1}^{\min(M_2-N_1, N_2)} V^{p}_{2i}X^{p}_{2i}
\end{align}
Here $X^{c}_{11}$, $X^{c}_{12}$, $\cdots$, $X^{c}_{1M_1}$ are independent Gaussian codewords from unit power codebooks that can be decoded by both receivers. Each $X^{c}_{1l}$ carries $\tfrac{N_1-\beta_{21}\min(M_1-N_2, N_1)}{M_1}$ DoF. $X^{p}_{11}$, $\cdots$, $X^{p}_{1\min(M_1-N_2, N_1)}$ and $X^{p}_{21}$, $\cdots$, $X^{p}_{2\min(M_2-N_1, N_2)}$ are independent Gaussian codewords from unit power codebooks that are intended to be decoded only by their desired receiver. Each $X^{p}_{1k}$ and $X^{p}_{2i}$ carries $\beta_{21}$ and $\bar{\beta}_{12}$ DoF, respectively.
$c_o$, $c_1$ and $c_2$ are scaling factors, $O(1)$ in $P$, chosen to ensure that the transmit power constraint is satisfied.

Here $V^{c}_{1l}$ are $M_1\times1$ generic unit vectors. $V^{p}_{1k}$ and $V^{p}_{2i}$ are $M_1\times1$ and $M_2\times1$ unit vectors, respectively, chosen so that
\begin{align}
\hat{\mathbf{H}}_{21}\begin{bmatrix}
V^{p}_{11}  & V^{p}_{12}   & \cdots & V^{p}_{1\min(M_1-N_2, N_1)}  
\end{bmatrix}={\bf 0}\\
\hat{\mathbf{H}}_{12}\begin{bmatrix}
V^{p}_{21}  & V^{p}_{22}   & \cdots & V^{p}_{2\min(M_2-N_1, N_2)}  
\end{bmatrix}={\bf 0}
\end{align}

The received signals are 
\begin{align}
Y_1=&c_o\sqrt{P}{\bf H}_{11}\sum_{l=1}^{M_1} V^{c}_{1l}X^{c}_{1l}+c_1\sqrt{P^{\beta_{21}}}{\bf H}_{11}\sum_{k=1}^{\min(M_1-N_2, N_1)} V^{p}_{1k}X^{p}_{1k}\nonumber\\
&+c_2\sqrt{P^{\bar{\beta}_{12}}}(\hat{\bf H}_{12}+\sqrt{P^{-\beta_{12}}}\tilde{\bf H}_{12})\sum_{i=1}^{\min(M_2-N_1, N_2)} V^{p}_{2i}X^{p}_{2i}+Z_1\\
=&c_o\sqrt{P}{\bf H}_{11}\sum_{l=1}^{M_1} V^{c}_{1l}X^{c}_{1l}+c_1\sqrt{P^{\beta_{21}}}{\bf H}_{11}\sum_{k=1}^{\min(M_1-N_2, N_1)} V^{p}_{1k}X^{p}_{1k}+O(1)+Z_1\\
Y_2=&c_o\sqrt{P}{\bf H}_{21}\sum_{l=1}^{M_1} V^{c}_{1l}X^{c}_{1l}+c_2\sqrt{P^{\bar{\beta}_{12}}}{\bf H}_{22}\sum_{i=1}^{\min(M_2-N_1, N_2)} V^{p}_{2i}X^{p}_{2i}\nonumber\\
&+c_1\sqrt{P^{\beta_{21}}}(\hat{\bf H}_{21}+\sqrt{P^{-\beta_{21}}}\tilde{\bf H}_{21})\sum_{k=1}^{\min(M_1-N_2, N_1)} V^{p}_{1k}X^{p}_{1k}+Z_2\\
=&c_o\sqrt{P}{\bf H}_{21}\sum_{l=1}^{M_1} V^{c}_{1l}X^{c}_{1l}+c_2\sqrt{P^{\bar{\beta}_{12}}}{\bf H}_{22}\sum_{i=1}^{\min(M_2-N_1, N_2)} V^{p}_{2i}X^{p}_{2i}+O(1)+Z_2
\end{align}

At Receiver 1 the multiplexed streams from Transmitter 1 are jointly decoded as a MAC channel. 
At the same time, at Receiver 2, $M_1$ elevated streams from Transmitter 1 can be decoded as a MAC channel (see Lemma \ref{lemma:mac} in the Appendix) by treating the remaining $\min(M_2-N_1, N_2)$ streams carrying $X^{p}_{2i}$ as white noise. 
After $X^{c}_{1l}$ are decoded and removed, all the rest of the signals can then be decoded separately along $\min(M_2-N_1, N_2)$ interference-free dimensions.

On the other hand, if $d_2=\min(M_2,N_2)=N_2$, then the achievability of $d_1=\min^+[\beta_{21}\min(M_1{-}N_2, N_1),$ $N_1-N_2+\beta_{12}\min(M_2{-}N_1, N_2)]$ can be shown similarly by simply switching the indices of the scheme in this subsection. The only difference is that when $N_1-N_2+\beta_{12}\min(M_2{-}N_1, N_2)\le0$, $\bar{\beta}_{21}\le0$, then in this case $d_1=0$. 

\section{Conclusion}
We studied the two-user MIMO interference channel with partial CSIT and arbitrary antenna configuration at each node. Through various examples we introduced the ideas of signal space partitioning and elevated multiplexing, and how they work together. Remarkably,  we found that there is a DoF benefit from increasing the number of  antennas at  a transmitter even if it has no CSIT and it already has more antennas than its desired receiver. Building upon these insights, a general achievable DoF result with partial CSIT was  presented. Generalizations of this work to the DoF region, more than $2$ users, and other settings besides interference networks, and including diversity of channel strengths are of immediate interest.

\appendix
\section*{Appendix}
Consider a multiple access channel with $K$ signal antenna transmitters. The receiver has $M$ antennas. The $M\times 1$ received signal vector $Y$ is represented as follows
\begin{align}
{\bf Y}=\sqrt{P}\sum_{k=1}^K {\bf H}_kX_k+\sum_{m=1}^M \sqrt{P^{\alpha_m}}{\bf G}_mZ_m
\end{align}
Here, $X_1, X_2, \cdots, X_K$ the transmitted symbols, normalized to unit transmit power constraint. $Z_m$ are i.i.d. Gaussian zero mean unit variance terms. The ${\bf H}_k, {\bf G}_n$ are $M\times 1$ generic vectors, i.e.,  generated from continuous distributions with bounded density, so that any $M$ of them are linearly independent almost surely. All $\alpha_m\in[0,1]$.

\begin{lemma}\label{lemma:mac}
The DoF tuple $(d_1, d_2, \cdots, d_K)$ is achievable in the multiple access channel described above, if
\begin{equation}
\sum_{i\in \max,k} d_i+\sum_{j\in \min,\min(k, M)} \alpha_j\le\min(k, M), \ \ \ \ \forall k\in[1, 2, \cdots, K]
\end{equation}
where $\sum_{i\in \max,k} d_i$ is the sum of the $k$ largest terms in $\{d_1, d_2, \cdots, d_K\}$ and $\sum_{j\in \min,\min(k, M)} \alpha_j$ is the sum of the $\min(k, M)$ smallest terms in $\{\alpha_1, \alpha_2, \cdots, \alpha_M\}$.
\end{lemma}

\proof  Choose all $X_i$ as zero mean unit variance i.i.d. Gaussians. A rate tuple $(R_1, R_2, \cdots, R_K)$ is achievable if the following inequalities are satisfied.
\begin{align}
\sum_{i\in\mathcal{U}} R_i\le I(\{X_i, \forall i\in\mathcal{U}\}; {\bf Y}|\{X_j, \forall j\in\mathcal{U}^c\}), \ \ \ \forall\ \mathcal{U}\subseteq\mathcal{K}, \label{region}
\end{align}
where $\mathcal{K}\define\{1, 2, \cdots, K\}$ and $\mathcal{U}$ can be any subset of $\mathcal{K}$, $\mathcal{U}^c$ is the complementary set of $\mathcal{U}$. 
\begin{align}
I&(\{X_i, \forall i\in\mathcal{U}\}; {\bf Y}|\{X_j, \forall j\in\mathcal{U}^c\})\nonumber\\
&=h({\bf Y}|\{X_j, \forall j\in\mathcal{U}^c\})-h({\bf Y}|\{X_j, \forall j\in\mathcal{K}\})\\
&=\min(|\mathcal{U}|, M) \log P+\sum_{j\in \max,M-\min(|\mathcal{U}|, M)} \alpha_j\log P-\sum_{j=1}^M\alpha_j\log P+o(\log P)\label{lemma3}\\
&=\min(|\mathcal{U}|, M) \log P-\sum_{j\in \min,\min(|\mathcal{U}|,M)} \alpha_j\log P+o(\log P)\label{1}
\end{align}
$|\mathcal{U}|$ is the cardinality of $\mathcal{U}$. $\sum_{j\in \max,M-\min(|\mathcal{U}|, M)} \alpha_j$ in \eqref{lemma3} is the sum of the $M-\min(|\mathcal{U}|, M)$ largest terms in $\{\alpha_1, \alpha_2, \cdots, \alpha_M\}$. $\sum_{j\in \min,\min(|\mathcal{U}|,M)} \alpha_j$ in \eqref{1} is the sum of the $\min(|\mathcal{U}|, M)$ smallest terms in $\{\alpha_1, \alpha_2, \cdots, \alpha_M\}$. \eqref{lemma3} follows from  Lemma 3 in \cite{geng_sun_jafar_ITW}.

From \eqref{1}, we  obtain the achievable DoF region,
\begin{align}
\sum_{i\in\mathcal{U}} d_i\le\min(|\mathcal{U}|, M)-\sum_{j\in \min,\min(|\mathcal{U}|,M)} \alpha_j, \ \ \ \forall\ \mathcal{U}\subseteq\mathcal{K}.
\end{align}
This concludes the proof.\hfill\QED

\bibliography{IEEEabrv,MIMOICpartial_full_rev}

\end{document}